\documentclass[prl,twocolumn,superscriptaddress]{revtex4}

\usepackage{times}
\usepackage{amsmath}
\usepackage{amssymb}
\usepackage{graphicx}
\usepackage{dcolumn}
\usepackage{bm}

\begin{document}


\title{The Effect of Landau Level-Mixing on the Effective
Interaction between Electrons in the fractional quantum
Hall regime}

\author{Waheb Bishara}
\affiliation{Department of Physics, California Institute of
  Technology, MC 256-80 Pasadena, CA 91125}
\author{Chetan Nayak}
\affiliation{Microsoft Research, Station Q, CNSI Building,
University of California, Santa Barbara, CA 93106-4030}
\affiliation{Department of Physics,
University of California, Santa Barbara, CA 93106}

\date{June 13, 2009}

\begin{abstract}
We compute the effect of Landau-level-mixing
on the effective two-body and three-body pseudopotentials
for electrons in the lowest and second Landau levels.
We find that the resulting effective three-body interaction
is attractive in the lowest relative angular momentum
channel. The renormalization of the two-body pseudopotentials
also shows interesting structure.
We comment on the implications for the $\nu=5/2$
fractional quantum Hall state.
\end{abstract}

\maketitle


\paragraph{Introduction.}
Deep in the quantum Hall regime, when
$\hbar {\omega_c} = \hbar e B/mc$ is the largest
energy scale, the Hamiltonian of a
two-dimensional electron liquid can be projected
into a single Landau level (LL), with all other LLs
being either completely filled or completely empty.
The projected Hamiltonian is often
tractable by numerical diagonalization
\cite{Yoshioka83,Laughlin83a,Laughlin83b,Haldane85a,Haldane85b}
(or, more recently, the density-matrix renormalization
group \cite{Shibata01,Feiguin08}).
Since the early days of the fractional
quantum Hall effect,
such numerical studies have
played a major role in establishing the viability of
various theories of observed quantum Hall plateaus.

The neglect of completely filled and empty LLs
is a good approximation so long as the Coulomb interaction scale
${e^2}/4\pi \epsilon {\ell_0}=({e^2}/4\pi\epsilon)\sqrt{eB/\hbar c}$
is much smaller than $\hbar {\omega_c}\equiv\hbar e B/mc$.
Corrections due to Landau-level-mixing
can be computed in powers of
$\kappa\equiv ({e^2}/4\pi\epsilon {\ell_0})/\hbar {\omega_c}\propto1/\sqrt{B}$
\cite{Sondhi92,Murthy02}.
Experiments in GaAs heterostructures and quantum wells
are done at magnetic fields $\sim 2-15\text{T}$, where
fractional quantum Hall states such as $\nu=1/3$
and $\nu=5/2$ are typically observed, which implies that
$\kappa \sim 0.6 - 1.8$. Thus, one might expect that the
effects of Landau-level-mixing would not be small.
Furthermore, even if small, Landau-level-mixing
(along with disorder) is the leading effect
which explicitly breaks particle-hole symmetry within
a LL. This may be of particular importance
in the second LL. The $\nu=5/2$ state
\cite{Willett87,Pan99,Eisenstein02,Xia04}
may be {\it inter alia} one of two candidates connected
by particle-hole conjugation, the Moore-Read Pfaffian
\cite{Moore91,Greiter92,Nayak96c} state and the anti-Pfaffian state
\cite{LeeSS07,Levin07}. In addition, a plateau is observed
at $\nu=12/5$ but not at the particle-hole conjugated
fraction, $\nu=13/5$ \cite{Xia04}. Numerical studies
of both $5/2$ \cite{Morf98,Rezayi00,Morf02,Feiguin08}
and $12/5$ have, thus far, neglected the
effects of LL mixing and may, thus, be questioned.

Thus, in this paper, we compute the renormalization
of the effective interaction within a LL,
due to virtual excitations of electrons into higher LLs
and holes into lower ones. We focus on the effective two-
and three-body interactions which are generated in
this way at lowest order in $\kappa$.
However, our analysis can be generalized in a
straightforward way to four-body and higher interactions
which are generated at higher order in $\kappa$.
We present our results in terms of Haldane's two-body
pseudopotentials \cite{Haldane85a} and the generalization to three-body
and higher pseudopotentials due to Simon {\it et al.}
\cite{Simon07}, which are the projections of the interaction onto
states of fixed relative angular momentum. Our results
can be used as an input for future numerical studies
of states in both the lowest and second LLs.
Our results have several interesting features. We find
that the effective interactions due to LL mixing are,
at lowest order, $\propto \kappa \cdot({e^2}/4\pi \epsilon {\ell_0})$
with a coefficient which is small. Thus,
even though the expansion parameter
$\kappa$ is O(1), a perturbative expansion may be valid.
In the lowest angular momentum channel relevant
to spin-polarized electrons,
the LL mixing contribution to the three-body
pseudopotentials is negative, i.e. is attractive.

\paragraph{Diagrammatics.}
We begin with the action:
\begin{multline}
S = \int \frac{d\omega}{2\pi} \sum_{m,n} \overline{c}^\alpha_{m,n}(\omega)
(i\omega - E_n + \mu)c^\alpha_{m,n}(\omega)\\
 -\,\frac{1}{2}\int \prod_{i=0}^4 \frac{d\omega_i}{2\pi}\, 
 \frac{e^2}{4\pi\epsilon \ell_0}V_{4,3;2,1} \,\,
 \overline{c}^\alpha_{m_4,n_4}(\omega_4)
 \overline{c}^\beta_{m_3,n_3}(\omega_3)\\
 \times\, c^\beta_{m_2,n_2}(\omega_2)c^\alpha_{m_1,n_1}(\omega_1)\,
 \delta^\omega_{43,21}
\end{multline}
where $c_{m,n}$, $\overline{c}_{m,n}$ are Grassman variables,
the $m_i$s are an orbital indices distinguishing states
within a LL and the $n_i$s are LL indices.
The spin indices $\alpha, \beta=\uparrow, \downarrow$ are summed
over whenever repeated.
$\delta^\omega_{43,21}$ is shorthand for
$2\pi\delta(\omega_4+\omega_3-\omega_2-\omega_1)$
and $V_{4,3;2,1}$ is shorthand for
$V({m_4},{n_4};{m_3},{n_3};{m_2},{n_2};{m_1},{n_1})$.
The single-particle energies are
$E_n=\hbar \omega_c \left(n+\frac{1}{2}\right)$,
and $\mu$ is the chemical potential, which we assume to
be equal to $\hbar {\omega_c}(N+1/2)$. Changes in
the chemical potential $\sim {e^2}/{4\pi\epsilon {\ell_0}}$,
which change the fractional filling of the $N^{\rm th}$
LL, do not affect the renormalization of the
effective interaction to lowest order in $\kappa$.
We have set the Zeeman energy to zero since it is much
smaller than either the Coulomb or cyclotron energies.
Spontaneous spin polarization in the $N^{\rm th}$ LL is not
precluded by this approximation and, in any case,  it is straightforward
to restore the Zeeman energy. The Coulomb interaction matrix elements
are given by the following expression, where
$G_{a,b}(q)$, for $a>b$, is \cite{MacDonald_LesHouches}
$G_{a,b}(q)=\left({b!}/{a!}\right)\left({-iq}/{\sqrt{2}}\right)^{a-b}L_b^{a-b}({|q|^2}/{2})$
where $L_b^{a-b}(q)$ is the generalized Laguerre polynomial:
\begin{multline}
V_{4,3;2,1}= 
\int \frac{d^2 q}{(2\pi)^2} \frac{2\pi}{|q|}  e^{-q^2}G_{n4,n1}(q^*)G_{n3,n2}(-q^*)
\\ \times\,[G_{m4,m1}(-q)]^2 .
\end{multline}
We assume that the $N^{\rm th}$ LL
(specializing later to $N= 0, 1$ LLL and SLL) is partially filled,
and integrate out all higher and lower LLs
to obtain the effective action $S_{\rm eff}$
in the $N^{\rm th}$ LL:
\begin{equation}
\label{eq:effective-action}
e^{iS_{\rm eff}[\overline{c}^{\alpha}_{m,N}, c^{\alpha}_{m,N}]}=
\int \prod_{n\neq N}\prod_{m}
[d\overline{c}_{m,n}(\omega)dc_{m,n}(\omega)]\, e^{iS}
\end{equation}
We carry out this integration perturbatively in the
Coulomb interaction which, as we will see, amounts
to an expansion in $\kappa$.
To first-order in this expansion parameter, $S_{\rm eff}$
will take the form:
\begin{multline}
\label{eqn:S-first-order}
S = \int \frac{d\omega}{2\pi} \sum_{m} \overline{c}^\alpha_{m,N}(\omega)
(i\omega - E_N + \mu)c^\alpha_{m,N}(\omega)\\
-\,\frac{1}{2}\int \prod_{i=0}^4 \frac{d\omega_i}{2\pi}\, 
 \frac{e^2}{4\pi\epsilon \ell_0}
 {u_2}({m_4},{m_3},{m_2},{m_1}) \,\delta^\omega_{43,21} \\
\times\, \overline{c}^\alpha_{m_4,N}(\omega_4)
\overline{c}^\beta_{m_3,N}(\omega_3)
  c^\beta_{m_2,N}(\omega_2)c^\alpha_{m_1,N}(\omega_1)\\
 -\,\frac{1}{3!}\int \prod_{i=0}^6 \frac{d\omega_i}{2\pi}\, 
 \frac{e^2}{4\pi\epsilon \ell_0}{u_3}({m_6},{m_5},{m_4},{m_3},{m_2},{m_1}) \\
\times\,\delta^\omega_{654,321} \,  \overline{c}^\alpha_{m_6,N}(\omega_6)
 \overline{c}^\beta_{m_5,N}(\omega_5)\overline{c}^\gamma_{m_4,N}(\omega_4)\\
 \times\, c^\gamma_{m_3,N}(\omega_3)c^\beta_{m_2,N}(\omega_2)
 c^\alpha_{m_1,N}(\omega_1)
\end{multline}
At zeroeth order, ${u_2}({m_4},{m_3},{m_2},{m_1})$
is simply the bare Coulomb interaction projected into
the $N^{\rm th}$ LL,
$V({m_4},N;{m_3},N;{m_2},N;{m_1},N)$. This is renormalized
at one loop, as we discuss below.
Meanwhile, ${u_3}({m_6},{m_5},{m_4},{m_3},{m_2},{m_1})$
is not present at zeroeth order and is only generated when
we take into account virtual transitions into higher LLs.
At higher orders, four-body, five-body, etc. interactions
${u_4}, {u_5},\ldots$ will be generated, but we do not discuss
these here.

\begin{figure}[t!]
\begin{center}
  \includegraphics[width=3.25in]{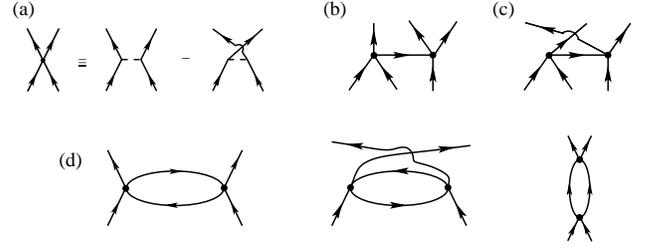}
  \caption{(a) The solid directed lines are electron lines
  and the dotted line denotes the Coulomb interaction.
  We define the four-point vertex ${\tilde V}_{4,3;2,1}^{\alpha'\beta';\beta\alpha}
  =V_{4,3;2,1}\delta^{\alpha\alpha'} \delta^{\beta\beta'}-
  V_{3,4;2,1}\delta^{\alpha\beta'} \delta^{\beta\alpha'}$ on the left-hand-side
  because the combination on
  the right-hand-side enters all of the diagrams which renormalize
  ${u_2}$ and ${u_3}$.
  (b) and (c) Two of the 9 diagrams which renormalize ${u_3}$.
  (d) The three diagrams which renormalize ${u_2}$.}
  \label{fig:diagrammatics}
\end{center}
\end{figure}

Our perturbative calculation is facilitated by the use of Feynman
diagrams, obeying the following rules.
We will always be considering diagrams in which
the Coulomb interaction enters in the combination
${\tilde V}_{4,3;2,1}^{\alpha'\beta';\beta\alpha}=
V_{4,3;2,1}\delta^{\alpha\alpha'} \delta^{\beta\beta'}-V_{3,4;2,1}
\delta^{\alpha\beta'} \delta^{\beta\alpha'}$ shown
in Fig. \ref{fig:diagrammatics}a. If we define a single
vertex as the difference between the two vertices
shown in Fig. \ref{fig:diagrammatics}a, then
there is only a single vertex in our calculations,
with two electron lines going in and two coming out.
It receives the vertex factor
${\tilde V}_{4,3;2,1}^{\alpha'\beta';\beta\alpha} \delta^\omega_{43,21}$,
where $m_1$, $m_2$ are the orbital indices of the
incoming electrons and ${m_3}$, ${m_4}$ are the orbital
indices of the outgoing electrons. In the diagrams of interest
to us, all but one or two of the particles' LL indices
$n_i$ will be that of the partially-filled LL, $N$, while
the internal legs will have LL indices which
range freely over $n\neq N$.
Every internal electron line receives the factor
$\frac{1}{i\omega-E_n}$,
where $\omega$ is the frequency of the line and
${E_n}=(n+\frac{1}{2})\hbar {\omega_c}$.
The lowest order diagrams with four and six external lines,
respectively, are in Fig. \ref{fig:diagrammatics}b-d.
To compute the effective interaction in the $N^{\rm th}$
LL, we require that all external legs in these
diagrams have $n_{\rm ext}=N$. The internal legs range over
all $n\neq N$. For $N=1$, this includes both virtual transitions
of holes to the LLL and of electrons to $N\geq 2$ LLs.

The diagrammatic expansion of this problem is similar
to that of a Fermi liquid, with the exception that momentum
integrals are replaced by sums over orbitals and Landau-level
indices. This has an important effect, namely that phase space
restrictions are much less severe. To see this,
consider the lowest-order diagram, shown in Fig.
\ref{fig:diagrammatics}b contributing to
the three-body effective interaction.
It corresponds to the following expression:
\begin{equation}
\label{eq:3par2order}
{\bigl(\mbox{$\frac{e^2}{4\pi\epsilon \ell_0}$}\bigr)^2}{\hskip -0.3 cm}
 \sum_{m_x; n_x\neq N}\!
\int \frac{d\omega_x}{2\pi}  \,
\frac{{\tilde V}_{6,x;2,1}^{\alpha'\lambda;\beta\alpha} \,
{\tilde V}_{5,4;x,3}^{\beta'\gamma';\gamma\lambda}}{i\omega_x
- (E_{n_x}-\mu)}
\delta^\omega_{6x,32}\,\delta^\omega_{54,x1}
\end{equation}
The energy $E_{n_x}-\mu$ is approximately
$\hbar \omega_c (n_x-N)$. As a result of the $\delta$-functions,
the $\omega_x$ integral in this particular diagram simply enforces energy conservation.
Note that there is a sum over LLs and orbitals, even though this
is a tree level diagram. In a zero-field Fermi liquid, by contrast,
there would be no momentum integral in such a diagram since the internal momentum would be completely fixed by the external ones.
From the energy denominator, we see that the contribution from virtual holes
in lower LLs is, naively, positive while the contribution from
virtual electrons in higher LLs is, naively, negative. However,
the matrix elements are not strictly positive, so the situation
is more complicated, as we will see in Table I.

In fact, there are $9$ diagrams similar to
Fig. \ref{fig:diagrammatics}b which
can be obtained from this one by permuting the external legs.
Another one of them is shown in
Fig \ref{fig:diagrammatics}c.
Note that these $9$ diagrams are actually $36$ diagrams when each vertex
is expanded as in Fig. \ref{fig:diagrammatics}a. These can be divided
into $6$ sets of $6$ diagrams, one set for each possible spin index structure
$\delta^{\pi(\alpha)\alpha'} \delta^{\pi(\beta)\beta'}\delta^{\pi(\gamma)\gamma'}$,
where $\pi$ is one of the $6$ permutations of $\alpha$, $\beta$, $\gamma$.
Summing them all up antisymmetrizes ${u_3}(m_6,...,m_1)\delta^{\alpha\alpha'} \delta^{\beta\beta'}\delta^{\gamma\gamma'}$
under all permutations of $\{({m_1},\alpha); ({m_2},\beta); ({m_3},\gamma)\}$
and of $\{({m_4},\gamma'); ({m_5},\beta'); ({m_6},\alpha')\}$.

Performing the integral in (\ref{eq:3par2order}), we find that
the resulting three-body coupling in the effective Lagrangian is given
by $-{u_3}(m_6,...,m_1)\,\delta^\omega_{654,321}
c^\dagger_{6}c^\dagger_{5}c^\dagger_{4}c_{3}c_{2}c_{1}$
(there is a minus sign since the contribution from (\ref{eq:3par2order})
has been re-exponentiated and combined with $-S$) with
${u_3}$ given by
\begin{multline} 
\label{eq:u654321}
{\sum_{\pi\in S_3}} (-1)^\pi {u_3}(\pi({m_6}),\pi({m_5}),\pi({m_4}),{m_3},{m_2},{m_1})\\
\times\,
\delta^{\alpha\pi(\alpha')} \delta^{\beta\pi(\beta')}\delta^{\gamma\pi(\gamma')} =\\
-\kappa \sum_{\omega,\theta'\in {C_3}}\left\{ \sum_{m_x; n_x\neq N}
\frac{ {\tilde V}_{\theta(6),x;\omega(2),\omega(1)}^{\theta(\alpha')\lambda;
\omega(\beta)\omega(\alpha)} \,
{\tilde V}_{\theta(5),\theta'(4);x,\omega(3)}^{\theta(\beta')\theta(\gamma');
\omega(\gamma)\lambda} }{{n_x}-N}\right\}
  \end{multline}
On the left-hand-side, we sum over permutations $\pi$
of $\{({m_4},\gamma'); ({m_5},\beta'); ({m_6},\alpha')\}$;
on the right-hand-side, we sum over cyclic permutations
of $\{({m_4},\gamma'); ({m_5},\beta'); ({m_6},\alpha')\}$
and of $\{({m_1},\alpha); ({m_2},\beta); ({m_3},\gamma)\}$
because the $4$-point vertex ${\tilde V}_{4,3;2,1}^{\alpha'\beta';\beta\alpha}$
has already been anti-symmetrized. In this expression, we have approximated the
energy denominator $i(\omega_3+\omega_2-\omega_6)-(E_{n_x}-\mu)$
by $-(E_{n_x}-\mu)=-\hbar {\omega_c}({n_x}-N)$.
The full expression gives an effective action
$S_{\rm eff}$ in (\ref{eq:effective-action}) with a frequency-dependent
${u_3}$, i.e. an interaction which is retarded
on time scales shorter than $1/(E_{n_x}-\mu)\propto 1/\hbar\omega_c$.
So long as all energies are small compared to $(E_{n_x}-\mu)$,
we can neglect these retardation effects by making this approximation.
This enables us to pass from the action (\ref{eqn:S-first-order}) to a
Hamiltonian.

The effective two-body interaction is
also renormalized at the same order,
so this contribution must also be kept for the sake of consistency (and may be equally
important for determining the ground state).  There are three
second-order diagrams contributing
to the renormalization of the two-body interaction,
which are familiar from Fermi liquid theory
(where Shankar has dubbed them the ZS, ZS', and BCS \cite{Shankar94}). They are depicted in Fig. \ref{fig:diagrammatics}d.
They give a contribution (one frequency integral has been performed
in each term in this expression):
\begin{multline}
\label{eq:2-body-renorm}
d{u_2}({m_4},{m_3},{m_2},{m_1})\delta^{\alpha\alpha'} \delta^{\beta\beta'}
-d{u_2}({m_3},{m_4},{m_2},{m_1})\delta^{\alpha\beta'} \delta^{\beta\alpha'}\\
 = 
\sum_{m_{x},m_{x  '}}\sum_{n_{x},n_{x'}\neq N}
\frac{{\tilde V}_{4,x';x,1}^{\alpha'\gamma';\gamma\alpha}
\, {\tilde V}_{x,3;2,x'}^{\gamma\beta';\beta\gamma'} \,
\bigl(\theta(\tilde{E}_{n_x})-\theta(\tilde{E}_{n_x'})\bigr)}{
\left(i({\omega_1}-{\omega_4}) + E_{n_x}-E_{n_x'})\right)}\\
- \sum_{m_{x},m_{x'}}\sum_{n_{x},n_{x'}\neq N}
\frac{{\tilde V}_{3,x';x,1}^{\beta'\gamma';\gamma\alpha} \,
{\tilde V}_{x,4;2,x'}^{\gamma\alpha';\beta\gamma'}\,
\bigl(\theta(\tilde{E}_{n_x})-\theta(\tilde{E}_{n_x'})\bigr)}{
\left(i({\omega_1}-{\omega_3}) + E_{n_x}-E_{n_x'})\right)}\\
-\frac{1}{2}
 \sum_{m_{x},m_{x'}}\sum_{n_{x},n_{x'}\neq N}
\frac{{\tilde V}_{4,3;x,x'}^{\gamma\gamma';\beta\alpha} \,
{\tilde V}_{x,x';2,1}^{\alpha\beta;\gamma'\gamma}\,
\bigl(\theta(\tilde{E}_{n_x})-\theta(-\tilde{E}_{n_x'})\bigr)}{
\left(i({\omega_1}+{\omega_2}) - (\tilde{E}_{n_x}+\tilde{E}_{n_x'})\right)}
\end{multline}
where $\tilde{E}_{n_x}\equiv E_{n_x}-\mu$.

\paragraph{Pseudopotentials}
A useful way of representing the results of evaluating
(\ref{eq:u654321}) and (\ref{eq:2-body-renorm}) is through Haldane's
pseudopotentials \cite{Haldane85a} and their generalization to three-body (and higher)
interactions by Simon {\it et al.} \cite{Simon07}. The idea is to project the interaction onto states of fixed
relative angular momentum within the $N^{\rm th}$ LL:
\begin{multline*}
V^{(2)}_{M,S} \equiv
  \sum_{\{m_i\}}
\langle M, M_{CM},S,{S^z}|{m_3},\beta;{m_4},\alpha\rangle\,\times\\
\langle {m_1},\alpha;{m_2},\beta| M, M_{CM},S,{S^z}\rangle
\,{u_2}({m_4},{m_3},{m_2},{m_1})
\end{multline*}
Here, $| M, M_{CM},S,{S^z} \rangle$ is a two-electron eigenstate of relative angular momentum $M$, center-of-mass quantum number
$M_{CM}$, total spin $S$ and total spin $z$-component $S^z$.
It is unimportant what basis we use for the center of mass
wavefunctions since the interaction is translationally-invariant, so the matrix
element is independent of $M_{CM}$. By spin-rotational symmetry,
it is also independent $S^z$. 
For $M$ odd, Fermi statistics requires $S=1$;
this case is relevant to a fully-polarized system. 
For $M$ even, $S=0$; these pseudopotentials only play a role when
both spin species are present.

Generalizing the three-body pseudopotentials of Ref. \cite{Simon07}
slightly to include spin, we have:
\begin{multline*}
V^{(3)}_{M,S,q,q'} \equiv
  \sum_{\{m_i\}}
\langle M,M_{CM},S,{S^z},q'|{m_4},\gamma;{m_5},\beta;{m_6},\alpha\rangle\\
\times\,\langle {m_1},\alpha;{m_2},\beta;{m_3};\gamma| M,M_{CM},S,{S^z},q \rangle
 \,{u_3}({m_6},...,{m_1})
\end{multline*}
Here, $| M,M_{CM},S,{S^z},q \rangle$, are the three-electron
states of relative angular momentum $M$ and total spin $S$.
Unlike the two particle case, there may be more than one
such state, which we label by the index $q$. The
first few $M=3,5,6,7,8$ only have a single such state,
so this extra index is superfluous. For instance,
$|M, M_{CM}=0,S=3/2,{S^z}=3/2\rangle$ for $M=3,5$ can be expressed
in the $|{m_1},{m_2},{m_3}\rangle$ basis as
$|3,0,3/2,3/2 \rangle=|0,\uparrow;1,\uparrow;2,\uparrow\rangle$
and $|5,0,3/2,3/2 \rangle=(\sqrt{3}|0,\uparrow;1,\uparrow;4\uparrow\rangle
-2|0,\uparrow;2,\uparrow;3\uparrow\rangle)/\sqrt{7}$. Similarly,
$|1, 0,1/2,1/2\rangle=$ $|0,\uparrow;0,\downarrow;1,\uparrow\rangle$.
From (\ref{eq:u654321}) and (\ref{eq:2-body-renorm}),
we compute the two-body and three-body pseudopotentials.
Our results are displayed in Table I.

\paragraph{Discussion.}
Table I contain the main results of this paper.
The most salient features are the following.
(1) The effects of
LL mixing are smaller than naively expected. The
corrections to the pseudopotentials are proportional to $\kappa$,
as expected, but the slight surprise is that the coefficient of $\kappa$ is $<0.2$; indeed, those that don't involve reversed spins are $<0.02$.
This small dimensionless number results from phase space
restrictions (which are important, albeit less so
than in a zero-field Fermi liquid),
partial cancellation between different excited LLs, and
the oscillatory nature of the relevant matrix elements.
Continuing our calculation to order $\kappa^2$, the contribution
to $V^{(3)}_{3,3/2}$ from one-loop diagrams is
$\approx 0.006\,\kappa^2$ \cite{Bishara09}.
Thus, our calculation might be valid to larger $\kappa$
than naively expected. In Ref. \onlinecite{Murthy02},
LL-mixing was considered in the Hamiltonian
approach using a modified charge operator and the Zhang-Das Sarma potential.
They found a 5\% effect on gaps for $\kappa=1$, further evidence of
the smallness of these effects.
(2) The $m=3$ three-particle pseudopotential,
$V^{(3)}_{3,3/2}$, is negative. This is not an {\it a priori}
obvious result since our computation is essentially
an RG calculation ({\it not} a calculation of the ground state energy)
and the $\beta$ function can be positive or negative. 
The physical significance of our result follows from the
observation that
the MR Pfaffian is the exact ground
state of a Hamiltonian with $V^{(3)}_{3,3/2}>0$ and all other
three-body and all two-body pseudopotentials equal to zero
\cite{Greiter92,Simon07}. The anti-Pfaffian state 
is the exact ground state of the particle-hole conjugate
Hamiltonian, which has $V^{(3)}_{3,3/2}<0$
and non-zero two-body pseudopotentials.
As seen in a recent preprint \cite{Wang09},
negative $V^{(3)}_{3,3/2}$ also favors the
anti-Pfaffian state over the MR Pfaffian state
near the Coulomb point.
The quantum phase transition between the two states
is first-order \cite{Wang09}, so the effects of a small
symmetry-breaking term will be magnified.
Thus, the $V^{(3)}_{m,s}$s, though small, may have
a large effect. However, it is possible
for a different state to be lower in energy than either the
anti-Pfaffian or MR Pfaffian. Thus, even though $V^{(3)}_{3,3/2}$
is larger in magnitude in the $N=0$ LL than in the $N=1$ LL
(and still negative), this does not
necessarily imply that the anti-Pfaffian state is expected there, too.
Finally, as noted above, the order $\kappa^2$ contribution
to $V^{(3)}_{3,3/2}$ is positive ($+0.006\,\kappa^2$),
so it is possible that $V^{(3)}_{3,3/2}$ changes sign as
$\kappa$ is increased and a phase transition occurs.
\begin{table}[t!]
\begin{tabular}{c | c c }
$m$ &  $V^{(3)}_{m,3/2}$ (LLL)    &   $V^{(3)}_{m,3/2}$ (SLL) \\
\hline
$3$ & $-0.0181$ & $-0.0147$ \\
$5$ & $0.0033$  & $-0.0054$ \\
$6$ & $-0.0107$ & $-0.0099$\\
$7$ & $0.0059$  & $ 0.0005$ \\
$8$ & $-0.0048$ & $ -0.0009$ \\
\end{tabular}
\begin{tabular}{c | c c }
$m$ &  $V^{(3)}_{m,1/2}$ (LLL)    &   $V^{(3)}_{m,1/2}$ (SLL) \\
\hline
$1$ & $-0.0346$ & $-0.0324$ \\
$2$ & $-0.0541$  & $-0.0315$ \\
\end{tabular}
\vskip 0.5 cm
\begin{tabular}{c | c c }
$m$ &  $\delta V^{(2)}_{m,1}$ (LLL)    &   $\delta V^{(2)}_{m,1}$ (SLL) \\
\hline
$1$ & $-0.0053$ & $0.0042$ \\
$3$ & $-0.0004$  & $0.0023$ \\
\end{tabular} {\hskip 0.2 cm}
\begin{tabular}{c | c c }
$m$ &  $\delta V^{(2)}_{m,0}$ (LLL)    &   $\delta V^{(2)}_{m,0}$ (SLL) \\
\hline
$0$ & $-0.1032$ & $-0.0325$ \\
$2$ & $-0.0012$  & $-0.0174$ \\
$4$ & $-0.0002$ & $ -0.0034$\\
\end{tabular}
\caption{Lowest order $3$-particle pseudopotentials, $V^{(3)}_{m,s}$,
and $2$-particle pseudopotentials, $\delta V^{(2)}_{m,s}$,
for $N=0,1$ (LLL and SLL).
The pseudopotentials are in units of
$\kappa \frac{e^2}{4\pi\epsilon \ell_0}$.
The table on the top left contains the pseudopotentials
for three particles with
total spin $S=3/2$ (relevant, for instance,
to the case of fully spin-polarized electrons) while the
table on the top right is for three particles with $S=1/2$.
The table on the bottom left contains the pseudopotentials
for pairs of particles with
total spin $S=1$ (relevant, for instance,
to the case of fully spin-polarized electrons) while the
table on the bottom right is for $S=0$.}
\label{table:Vm3_lowest}
\end{table}
(3) It is important to consider the
higher $V^{(3)}_{m,3/2}$s, which oscillate
with $m$. It is also important to consider the $\delta V^{(2)}_{m,1}$s,
which have opposite signs in the $N=0$ and $N=1$ LLs.
Landau level mixing suppresses the ratio $V^{(2)}_{1,1}/V^{(2)}_{3,1}$
more strongly in the $N=1$ LL than in the $N=0$ LL. This may
help explain why a quantum Hall plateau does not develop in
the $N=0$ LL, unlike in the $N=1$ LL.
(4) The pseudopotentials which come into play when
the electrons in the $N^{\rm th}$ LL are not fully spin-polarized
are more strongly renormalized. Thus, apart from
particle-hole symmetry-breaking, the effects of
LL mixing may be strongest in partially-polarized
or unpolarized states.
At any rate, our results should be viewed
as an input for exact diagonalization and DMRG studies
of quantum Hall states: with the effects of Landau level mixing
incorporated in the starting Hamiltonian, one could thereby solve
a more realistic model.

\acknowledgements
We would like to thank P. Bonderson, S. Das Sarma,
A. Feiguin, S.-S. Lee, M. Levin, R. Morf, E. Rezayi, S. Ryu, K. Shtengel,
S. Simon, and Kun Yang for discussions. As we were completing this work,
we learned of related unpublished work by E. Rezayi and S. Simon
\cite{Rezayi} in which the quantum Hall Hamiltonian, truncated to
three spin-polarized Landau levels, is numerically diagonalized.

\end{document}